\documentclass[prb,amsmath,amssymb]{revtex4}
\usepackage{epsf}
\usepackage[T1]{fontenc}
\usepackage{calc}
\usepackage{bm}
\usepackage{setspace}

\begin{document}

\title{Black hole entropy and the holographic principle
\footnote{Contribution to the book {\it Advances in 
Physical Sciences}, ed. Luis D. Carlos, (Aveiro University Press, 2008), 
proceedings of the meeting 
held at the Universidade de Aveiro, September 2005.}}
\author{Jos\'e P. S. Lemos}
\affiliation{Centro Multidisciplinar de Astrof\'{\i}sica - CENTRA,
Departamento de F\'{\i}sica, Instituto Superior T\'ecnico - IST, 
Universidade T\'{e}cnica de Lisboa - UTL, Av.
Rovisco Pais 1, 1049-001 Lisboa, Portugal\,\,}
\email{lemos@fisica.ist.utl.pt}
\begin{abstract}
Black holes monopolize nowadays the center stage of fundamental
physics.  Yet, they are poorly understood objects.  Notwithstanding,
from their generic properties, one can infer important clues to what a
fundamental theory, a theory that includes gravitation and quantum
mechanics, should give. Here we review the classical properties of
black holes and their associated event horizons, as well as the
quantum and thermodynamic properties, such as the temperature, derived
from the Hawking radiation, and the entropy.  Then, using the black
hole properties we discuss a universal bound on the entropy for any
object, or for any given region of spacetime, and finally we present
the arguments, first given by 't Hooft, that, associating entropy with
the number of quantum degrees of freedom, i.e., the logarithm of
quantum states, via statistical physics, leads to the conclusion that
the degrees of freedom of a given region are in the area $A$ of the
region, rather than in its volume $V$ as na\"ively could be thought.
Surely, a fundamental theory has to take this in consideration.
\end{abstract}
\vspace{4cm}
\maketitle

\section{Introduction}

Black holes have been playing a fascinating role in the development of
physics. They have entered into the physics domain through a
combination of the disciplines of general relativity and astrophysics.
Indeed, black holes arise naturally within the theory of general
relativity, Einstein's geometric theory of gravitation.  Being exact
vacuum solutions of the theory, they are thus, unequivocally,
geometric objects. From the first solution in 1916, the Schwarzschild
black hole, to the rotating Kerr black hole solution found in 1963,
until they were accepted as the ultimate endpoints of the
gravitational collapse of massive stars, as well as the gravitational
collapse of huge amounts of matter (being it, clusters of stars, dark
matter or any other matter form) in the center of galaxies, there has
been a highly winding road (see, e.g., \cite{htww,penrose65,mtw} for
the initial developments and references therein).  The name black hole
was coined in 1968 by Wheeler \cite{wheelerbh} (see also
\cite{wheelerautobiog,bhname}).  Now there is no doubt that solar mass
black holes abound in our Galaxy, and supermassive ones reign at the
centers of galaxies playing their roles as energizers of their own
neighborhoods, such as in quasars (see \cite{kormendy} for the status
of supermassive black holes in galaxies).  New theoretical
developments show that black holes can form in various ways. They can
be eternal being out there since the very initial universe, they can
be pair created in a Schwinger type process, they can form from the
collision of highly energetic particles as in accelerators, and in the
more usually case, discussed in the original work that gave rise to
the concept, they can form from the collapse of matter.  These and
other developments, which have put black holes at the center of
studies in fundamental physics, were possible after Hawking discovered
that they act as thermal quantum creators and radiators of particles
\cite{hawking1975}. This result, albeit in a semiclassical regime,
unites in one stroke, gravitation (represented by the universal
constant of gravitation $G$, and the velocity of light $c$), quantum
mechanics (represented by Planck's constant $\hbar$) and statistical
physics (represented by Boltzmann's constant $k_{\rm B}$). Thus, black
holes turn out to be in the forefront of physics, since by acting as
unifying objects, through them one can test unifications ideas of
gravity (and possibly other fields) and quantum mechanics.

\section{Classical properties of black holes}

\subsection{The event horizon}
General relativity introduces the idea that gravitation is a
manifestation of the geometry and curvature of spacetime. Its
equations, Einstein's equations, imply that objects, like test
particles either massive or massless (like light), move as geodesics
in the given underlying curved geometry which, in turn, is established
by a certain concentration of matter and energy. Einstein's equations
imply in addition that a high concentration of matter and energy curve
spacetime strongly.  When the concentration of matter and energy is
high enough, such as in a collapsing star, spacetime will be so curved
that it tears itself, so to speak, and forms a black hole. And, like a
burst of water in a river, that suddenly opens up a falls, once
falling down the falls it is impossible to get back. Indeed, a black
hole is a region where the spacetime curvature is so strong that the
velocity required to escape from it is greater than the speed of
light. The surface of the black hole is the limit of the region from
where light cannot escape. Just outside this surface light and
particles may escape and be detected at infinity. Inside the surface
all particles fall through and never come out again.  This boundary,
defining what can be seen by observers outside the black hole, the
boundary of the region of no return to the outside, is called the
event horizon.  Thus, in the case the black hole forms from a
collapsing star, say, the event horizon is not to be identified with
the surface of the star that formed the black hole.  The matter that
formed the black hole goes through its own horizon, and once inside
the horizon, it will continue to collapse right down the falls it
created, until a singularity forms, where the curvature of spacetime
blows up, i.e., tidal forces disrupt spacetime itself. There is a
horizon floating outside, whereas the surface of the star and indeed
the whole of the star are now at the singularity.  In a black hole,
the singularity is hidden behind the horizon, in the hidden region,
where nothing comes out.  In brief, a black hole is not a solid body
with a matter surface, it is a three surface in space and time bounded
by a horizon. It is a pure gravitational object with an event horizon,
from inside which there is no escape, and with a hidden singularity at
its center.  A singularity is an object nobody knows what it is.  To
know what lies inside a black hole, what is a singularity, it is
certainly one of the most important problems to be solved in physics,
but due to its complexity one sees it seldom discussed. Here we also
do not discuss it. Rather, we are interested in the horizon and in the
black hole properties, classical and quantum, exterior to it.

In the simplest case the horizon is a sphere, and the corresponding
solution is called the Schwarzschild solution. In this case the
horizon radius $R_{\rm bh}$ scales with the mass $M$ of the black hole,
$R_{\rm bh}=2M\,\left(\frac{G}{c^2}\right)$, or in natural or Planck units
($G=1$, $c=1$, $\hbar=1$, $k_{\rm B}=1$,  
units which will be often used), one has
\begin{equation}
R_{\rm bh}=2M\,,
\label{horizonradius}
\end{equation} 
so that the more massive the black hole the larger the horizon 
(see, e.g., \cite{mtw}). One
can think of the horizon as a sphere of photons, or null geodesics,
that are trying to get out radially, but due to the strong
gravitational force, i.e., high curvature, stay fixed at
$R_{\rm bh}=2M$. Photon spheres inside the horizon,
in the hidden region, are dragged down to the singularity even if they
are locally outgoing. On the other hand, photon spheres outside the
horizon, moving radially, reach infinity, with those that originate in
the horizon's vicinity having to climb a huge gravitational field, or
geometrical barrier, take a long time to do so.

\subsection{Black holes have no hair}
In general relativity, time depends on the observer. Observers,
particles, whatever that enters the black hole will go straight to the
singularity, and will not come back, according to their own
reckoning. On the other hand, observers that stay outside the horizon,
see things differently. They cannot know what happens inside a black
hole. Classically, nothing, no signal nor information, can emerge
across the horizon and escape into the outside to be detected by an
observer in the external world. However, at least one feature of the
spherical black hole can be measured outside. One can know from the
outside the mass of the black hole. Thus, if the Schwarzschild black
hole was formed from the collapse of matter, the only property of the
initial matter that can be known after it has collapsed is its mass,
all the knowledge on the other properties of the initial matter has
disappeared down the horizon.  Moreover, additional matter falling
into the black hole simply adds to the mass of the black hole, and
disappears from sight, taking with itself its own properties. More
generally, within general relativity, there are three parameters that
can characterize a black hole.  A spherically symmetric vacuum black
hole, the Schwarzschild black hole, is characterized only by its mass
$M$, with the horizon radius given by equation
(\ref{horizonradius}). A great deal of complexity is added if in
addition to the mass $M$ the black hole possesses angular momentum
$J$, with the horizon being now oblate instead of spherical. This is
the important rotating Kerr black hole solution, an exact solution,
that by introducing new dynamics, gave a totally new flare to
classical black hole theory, which in turn was essential in the
construction of the coupling of black holes to quantum mechanics.
Adding electric charge $Q$ one has the Kerr-Newman black hole family,
where for $J=0$ one calls the charged nonrotating black hole a
Reissner-Nordstr\"om black hole. So, a generic Kerr-Newman black hole
is characterized three parameters only, namely $(M,J,Q)$, this being
the only knowledge one obtains out of a black hole (see, e.g.,
\cite{mtw}). Such a black hole can form from the collapse of an
extremely complex rotating distribution of ions, electrons, radiation,
all kinds of other matter, and myriads of other properties
characterizing the matter itself.  But once it has formed, for an
external observer, the only parameters one can know from the outside
are the the mass-energy $M$ of the matter that went in, its angular
momentum $J$, and its electrical charge $Q$. One then says a black
hole has no hair, since it has only three hairs, and someone with
three hairs is effectively bald (see, e.g., \cite{mtw}).  This
property has measurable consequences on the spacetime outside the
horizon.  The black hole's mass, angular momentum, and electric charge
certainly change accordingly to the type of matter that is added onto
it. Moreover, this change obeys strict rules, indeed, one can show
that the laws of conservation of energy, angular momentum and charge,
are still valid when a black hole is involved \cite{bardeenetal}.
Thus, $M$, $J$, and $Q$ are observable properties that can be known
through some form of external interaction with the black hole. On the
other hand, all the other features that could possibly further
characterize the black hole do not exist. Or if they exist they have
vanished from sight.  These other features certainly characterize
usual matter, star matter say. But once the star has imploded into a
black hole all the features, but three, disappear. Where are now those
features? Is this property of hiding features, one that one can
capitalize on, and discover new properties of the world? By a
remarkable twist, quantum mechanics comes to the rescue.

\section{Quantum properties of black holes
}

The questions raised lead us into the quantum realm and put black
holes on a central scene to unify in one stroke, gravitation and
quantum mechanics itself, within the framework of thermodynamics and
through the concepts of black hole temperature and entropy.  These
results will also definitely impel into an ultimate bound on the
entropy of a given region and to the the establishment of a new
revolutionary holographic principle. Let us see each point one at a
time, and then all altogether.

There are now many theories, of general relativity type, that have
many different black hole solutions, with charges other than $M$, $J$,
and $Q$. These theories, one way or the other have general relativity
as a limit.  So let us stick to general relativity, and moreover let
us study the simplest case the Schwarzschild black hole, with its only
one hair, the mass $M$ and the associated horizon radius $R$ given in
equation (\ref{horizonradius}).  The Kerr black hole was very
important to put nontrivial dynamics on a classical setting, it acted
in this context as a catalyst, but after it induced forcefully the
introduction of quantum phenomena in the whole scheme, one can use
the simplest black hole, the Schwarzschild black hole, this is indeed
sufficient to understand the profound ideas that lie underneath black
hole physics.

\subsection{Black hole thermodynamics}

The first law of thermodynamics states that the total energy is
conserved in an isolated system. It seems trivially obvious nowadays,
but some time ago it was hard to understand the nature of heat as
energy, an achievement that was accomplished after the work of Carnot
first, and the further insights of Mayer, Joule, Kelvin, and Helmholtz
in the 1840s \cite{lindley,brush1,brush2}. In its simple form it
states that $dE=dQ$, where $Q$ is the amount of heat exchanged, and
$dE$ is the variation in internal energy of the system.  Another
important idea in thermodynamics, and in physics in general, is the
introduction of the concept of $dQ/T$, heat over temperature, a state
variable.  It was devised by Clausius in 1854, who also found an
adequate name for it in 1865, entropy $S$, such that $dS=dQ/T$. In
terms of state variables the first law can now be written as
$dE=T\,dS$. This concept of entropy, also led Clausius to postulate a
second law of thermodynamics by stating that equilibrium states have
an entropy associated with them such that processes can occur only
when the final entropy is larger than the initial entropy, i.e., in
any closed system, entropy always increases or stays the same, $d
S\geq0$.  As also worked out by him in a paper dealing with the ``the
nature of the motion which we call heat'', the entropy concept had an
immediate impact in kinetic theory and statistical mechanics (see,
e.g., \cite{brush2}). Both these advances are remarkable. Of course,
Clausius could not know of its ramifications and problems that such a
concept would introduce more than 100 years later, when applied to
gravitating systems. A first hint of these problems appeared in
studies on Newtonian gravitating systems, such as in clusters of stars
\cite{lyndenbell1,lyndenbell2}, whereas one needs general relativity
to apply thermodynamic and statistical mechanics concepts to fully
general relativistic objects.  Two such systems are the Universe
itself and a black hole. That the entropy concept and its associated
second law of entropy increase can have remarkable implications upon
the Universe as a whole and on the arrow of time was first understood
by Boltzmann within his statistical physics formulation (see, e.g.,
\cite{brush2,eddington}), an issue that is still today under heavy
discussion \cite{huw,penrosebook}.  But conundrums of a different
caliber and with a more direct physical significance perhaps,
involving physics at the most fundamental level, have arisen from the
fact that a black hole has entropy, an entropy with a form never seen
before.  Indeed, black hole entropy is proportional to the area of the
black hole, rather than the volume. Let us first see how the black
hole entropy arises and then where the second law takes us to.

If one thinks, as before, of a black hole forming from the
collapse of a matter star, one has an initial configuration, a star
say, and a final configuration, a black hole. The star is specified by
very many parameters and quantities, the black hole by the mass $M$
alone, in the spherical vacuum case.  This led us to argue above that a
black hole is a system specified by one macroscopic hair parameter
only, the mass $M$ and hiding lots of other parameters perhaps located
inside the black hole event horizon.  Thus, the black hole acts like a
black box. In physics there is another instance of this kind of black
box situation, whereby a system is specified and usefully described by
few parameters, but on a closer look there are many more other
parameters that are not accounted for in the gross macroscopic
description. This is the well know case of thermodynamics described
above.  For thermodynamical systems in equilibrium one gives the
energy $E$, the volume $V$, and the number of particles $N$, say, and
one can describe the system in a useful manner, obtaining from the
laws of the thermodynamics its entropy and other important
quantities. On doing this one does not worry that the system encloses
a huge number of molecules and that the description hides its own
microscopic features.  Of course, one can then plunge into a deeper
treatment and apply statistical mechanical methods to the particles
constituting the thermodynamical system, using the distribution
density function of Gibbs for classical particles or the density
matrix for quantum ones in the appropriate ensemble, and then applying
Boltzmann's formula for the entropy $S=k_{\rm B}\,\ln\Omega$, where
$\Omega$ is the number of states, or any other formula, like
Gibbs' formula, to make the connection to thermodynamics.  Due to this
black box analogy between a black hole and a thermodynamic system, one
can ask first the question: Is thus a black hole a thermodynamic
system? If yes, one should pursue and ask two further questions: Can
one find the analogue of the constituent particles to allow for a
statistical interpretation?  To where can the black hole thermodynamic
system lead us to, in terms of the ultimate fundamental theory?

The first, and then the subsequent questions, started to be answered
through a combination of hints. From the Penrose and superradiance
processes, deduced using Kerr black hole backgrounds, 
one could conclude that the area of a black hole would not
decrease in such cases
\cite{penroseprocess,superradiancep1,superradiancep2,christodoulou},
an idea that culminated with the underlying area law theorem,
which states that in a broad class of circumstances, such as in black
hole merger events, the area could never decrease, only increase or
stay even in any process \cite{hawkingarealaw}. At about the same
time, Wheeler raised the problem (see \cite{wheelerautobiog}), that when
matter disappears into a black hole, its entropy is gone for good, and
the second law seems to be transcended, i.e., in the vicinity of a
black hole entropy can be dumped onto it, thus disappearing from the
outside world, and grossly violating the second law of thermodynamics.
Bekenstein, a Ph.D. student in Princeton at the time, solved part of
the problem in one stroke. With the hint that the black hole area
always increases, he postulated, entropy is area
\cite{bekensteins-a}. Specifically, he postulated
\cite{bekensteins-a}, $S_{\rm bh}=\eta\,\frac{A_{\rm bh}}{A_{\rm
pl}}\;\,k_{\rm B}\,, \label{entropybekenstein}$ where $A_{\rm bh}$
is the black hole area, $\eta$ is a number of the order of unity or
so, that could not be determined, $A_{\rm pl}$ is the Planck area,
and $k_{\rm B}$ is the Boltzmann constant.  Note that the Planck
length $l_{\rm pl}\equiv\sqrt{\frac{G\hbar}{c^2}}$, of the order of
$10^{-33}\,$cm, is the fundamental length scale related to gravity and
quantum mechanics, and the Planck area is its square, $A_{\rm
pl}=l_{\rm pl}^2\sim 10^{-66}\,{\rm cm}^2$. Several physical arguments
were invoked to why the entropy $S$ should go with $A_{\rm bh}$ and
not with $\sqrt {A_{\rm bh}}$ or $A_{\rm bh}^2$.  For instance, it
cannot go with $\sqrt {A_{\rm bh}}$. This is because 
$A_{\rm bh}$ itself goes
with 
$R_{\rm bh}^2\sim M^2$, for a Schwarzschild black hole, and when two black
holes of masses $M_1$ and $M_2$ merge, the final mass $M$ 
obeys $M<M_1+M_2$ since there is emission
of gravitational radiation.  But if $S_{\rm bh}\propto\sqrt {A_{\rm bh}}
\propto M<M_1+M_2\propto S_{\rm bh1}+S_{\rm bh2}$ the entropy could decrease,
so such a law is no good. The correct option turns out to be $S_{\rm
bh}\propto A_{\rm bh}$, the one that Bekenstein took.  It seems thus,
there is indeed a link between black holes and thermodynamics.  In
addition, it seems correct to understand that this phenomenum is a
manifestation of an underlying fundamental theory of spacetime, a
quantum theory of gravity, since the Planck area appears naturally in
the formula, hinting that there must be a connection with some
fundamental spacetime microscopic ingredient whose statistics connects
to the thermodynamics.

Since there is a link between black holes and thermodynamics, black
holes have entropy, one can then wonder whether black holes obey the
first and second laws of thermodynamics (see, e.g., 
\cite{parker,davies,lemos1} for reviews on black hole particle creation
and black hole thermodynamics). 
In relation to the first law,
note that for a Schwarzschild black hole, the simplest case, one has
that the area of the event horizon is given precisely by $A_{\rm
bh}=4\pi\,R_{\rm bh}^2$. Now, 
$R_{\rm bh}=2M$, so one has $A_{\rm bh}=16\pi M^2$ (in
natural or Planck units).  Then one finds $dM=1/(32\,\pi\,M)\, dA_{\rm
bh}$, which can be written as \cite{bardeenetal}, 
\begin{equation}
dM=\frac{\kappa}{8\,\pi}\, dA_{\rm bh}\,, 
\label{firstlawofbhs}
\end{equation} 
where $\kappa$ is the surface gravity of the black hole, a quantity
that can be calculated independently and gives a measure of the
acceleration of a particle at the event horizon.  In the Schwarzschild
case $\kappa=1/4\,M$.  Equation (\ref{firstlawofbhs}) is a simple
dynamical equation for the black hole. When one compares it with the
first law of thermodynamics,
$
dE=TdS\,,
\label{firstlawofthermodynamics}
$
the similarity is striking, and since following Bekenstein $S_{\rm bh}
$ and $A_{\rm bh}$ are linked, and following Einstein $M$ and $E$ are
linked, indeed they are the same quantity, one is tempted to associate
$T$ and $\kappa$ \cite{waldbook}.  But from thermodynamical arguments
alone one cannot determine $\eta$ the dimensionless proportionality
constant of order unity between entropy and area, and cannot also
determine the constant of proportionality between $T$ and $\kappa$,
related to $\eta$.  Using quantum field theory methods in curved
spacetime Hawking \cite{hawking1975}, in a spectacular tour de force,
showed that a Schwarzschild black hole radiates quantically as a black
body at temperature $T_{\rm bh}=\frac{1}{8\pi
M}\left(\frac{\hbar\,c^3}{G\,k_{\rm B}}\right)$, uniting in one
formula $\hbar$, $G$ and $c$, and $k_{\rm B}$.  In natural Planck
units, and returning to $\kappa$ this is,
\begin{equation}
T_{\rm bh}=\frac{\kappa}{2\pi}\,, 
\label{hawkingtemperature}
\end{equation}
connecting definitely and physically the surface gravity with
temperature, and closing the thermodynamic link.  Moreover, from the
first law of thermodynamics one obtains $\eta=1/4$, yielding finally
\begin{equation} 
S_{\rm bh}=\frac14\,A_{\rm bh}\,,
\label{entropyarealawgeometrical} 
\end{equation} 
in natural units. Thus, Hawking radiation allows one to determine, on
one hand, the relation between the temperature of the black hole and
its surface gravity, and on the other hand, to fix once and for all
the proportionality constant between black hole entropy and horizon
area.  The black hole entropy is one quarter of the event horizon's
area, when measured in Planck area units.  For thermodynamic systems,
this is a huge entropy, the entropy of a black hole one centimeter in
radius is about $10^{66}$ in Planck units, of the order of the
thermodynamic entropy of a cloud of water with $10^{-3}$ light years
in radius. The Hawking radiation solved definitely the thermodynamic
conundrum. The generalized first law is then given by a simple
extension of equation (\ref{firstlawofbhs}). $M$ is now the energy of
the whole system, black hole plus matter, $T$ for the matter and
$\kappa/2\pi$ for the black hole have the same values, for a system in
equilibrium, and the entropy of the thermodynamic system is now
$S=S_{\rm bh}+S_{\rm matter}$, a sum of the black hole entropy $S_{\rm
bh}$, and the usual entropy of the matter and radiation fields which
we denote simply as $S_{\rm matter}$.  It is advisable to separate the
entropy into two terms, since one does not know for sure the meaning
of black hole entropy.

What about the second law of thermodynamics, can it be embodied in a
framework where black holes are present?  The second law of
thermodynamics mathematizes the evidence that many processes in nature
are irreversible, hot coffee cools in the atmosphere, but cold coffee
never gets hot spontaneously, and so on. The law states that the
entropy of an isolated physical system never decreases, either remains
constant, or it increases, usually. It holds in a world where
gravitational physics is unimportant. What happens in gravitational
systems in which there are black holes.  Given that the black hole is
a thermodynamic system, with entropy and temperature well defined, the
second law of thermodynamics $d\,S\geq0$ should be obeyed.  Indeed one
can write the second law as
\begin{equation}
dS_{\rm bh}+d S_{\rm matter}\geq0\,,
\label{gsl}
\end{equation}
commonly called the generalized second law
\cite{bekensteingeneralizedsecondlaw}.  In words, the sum of the black
hole entropy and the ordinary entropy outside the black hole cannot
decrease. This generalized second law proved important in many
developments, and its consequences are the main object of this review.
The generalized second law has passed several tests.  For instance,
when a star collapses to form a black hole, one can show that the
black hole has an entropy that far exceeds the initial entropy of the
star. Also, when matter falls into an already existing black hole, the
increase in black hole entropy always compensates for the lost entropy
of the matter down the horizon.  Another interesting example where the
generalized second law holds involves Hawking radiation. Due to this
radiation the black hole evanesces. Its mass decreases, and so the
black hole area also decreases. This violates the area law theorem,
but this is no problem, the theorem was proved classically.  Then, the
black hole entropy decreases indeed. However, one can show that the
entropy in the emitted radiation exceeds by some amount the original
entropy of the black hole, upholding the generalized second law
\cite{sorkinwalzhang1981}.  Using generic arguments hinged on a
quantum definition of entropy, it is possible to argue, that due to
lack of influence of the inside on the outside, the generalized second
law is valid for processes involving black holes \cite{sorkin86}.  We
note that there are arguments that claim that one does not need the
generalized second law, the ordinary second law alone is enough in
itself, see, e.g., \cite{unruh82,bekenstein83,unruh83,bekenstein94}.
This controversy would merit a review in itself.

\subsection{Black hole entropy}

Before start discussing to where the generalized second law leads us,
it is interesting to think about the consequences of black holes
having entropy, as Bekenstein did almost immediately after his major
discovery \cite{bekensteins-a}.  Entropy is one of the most important
concepts in everyday physics.  Somehow, it is a recondite concept, and
even more mysterious when black holes are involved. Let us see this.

Following Boltzmann, the entropy $S$ of a closed isolated system with
fixed macroscopic parameters, is given by,
\begin{equation}
S=k_{\rm B}\ln\Omega\,,
\label{boltzentropy0}
\end{equation}
where again $k_{\rm B}$ is the Boltzmann constant, and $\Omega$ is the
number of accessible microstates that the large system has. Each
microstate $i$ has equal probability $p_i$ of occurring, so
$p_i=1/\Omega$, and equation (\ref{boltzentropy0}) can be written in
the alternative form $S=-k_{\rm B}\ln\,p_i$. For open systems, that
can exchange energy and other quantities, the entropy can be written
in a more useful manner as
$
S=-k_{\rm B}\sum_ip_i\ln p_i\,,
\label{gibbs}
$
where $p_i$ is the probability of microstate $i$ occurring, which now
due to the openness of the system is not anymore equal for each state,
states with a given energy, the average energy, have a higher
probability of occurring. This formula was given by Gibbs upon careful
consideration of his ensemble theory and generalization of Boltzmann
ideas (see, e.g., \cite{cowan} for the deduction of Gibbs entropy
formula from equation (\ref{boltzentropy0})).  If the system is closed
then $p_i=1/\Omega$ and Boltzmann equation (\ref{boltzentropy0})
follows.  In the Gibbs formulation, one works in a $6N$ dimensional
classical phase space, and having to work with a continuum
distribution probability density, the phase-space density $\rho$
(instead of $p_i$), one should write $S=-k_{\rm
B}\int\,d^{3N}q\,d^{3N}p\,\,\rho(q,p)\,\ln\rho(q,p)$, which is the
continuum Gibbs entropy equation for a system of $N$ particles in
three dimensional space with $6N$ classical degrees of freedom, $3N$
for the coordinates and $3N$ for the momenta. In this setting each
point in the phase-space represents a state, a microstate, of the
system.  This was then generalized, in a natural way, although through
a postulated basis, by Von Neumann to quantum systems. One postulates
first that $\rho$ goes into the quantum operator $\hat\rho$ which
gives the probability that the system is in some given microstate
(essentially is $p_i$), and second that the entropy is $S=-k_{\rm
B}{\rm Tr}\,\hat\rho\ln\hat\rho$. This Von Neumann entropy should be
calculated in some complete orthonormal basis of the appropriate state
space, or Hilbert space. Since $\hat\rho$ can have non-diagonal terms,
which can be suited for calculating quantities other than traces, the
Von Neumann entropy is a generalization of the Gibbs entropy, although
possibly not unique. All these formulas for the entropy can be useful,
depending on the context one is working.  Gibbs formula, for instance,
has an interesting advantage sometimes.  Indeed, the formula is the
same as the one that emerged for the entropy in information theory,
the Shannon entropy \cite{shannonbook}.  The Shannon entropy first
appeared in connection with a mathematical theory of communication,
where it was perceived that the best measure of information is
entropy. In fact, entropy in an informational context represents
missing information.  The Shannon entropy formula is given by
$S=-k_{\rm S}\sum_i p_i\ln p_i\,$, where here, since the connection
with temperature is unimportant, $k_{\rm B}$ is substituted by $k_{\rm
S}$, the Shannon constant, 
which generally is put equal to $1/\ln2$, so that the entropy is
given in bits, a dimensionless quantity.  Apart the constant used, 
$k_{\rm B}$ or $k_{\rm S}$, which is a matter of convenience, the two
entropies are the same. However, Shannon entropy is applied to measure
the information a given system (a computer for instance) has,
basically how many bits the system has, whereas Gibbs entropy is
applied to the thermodynamic system itself, essentially the number of
molecules the system (a computer for instance) has. Both entropies can
be given in Shannon units, of course. Gibbs entropy is usually much
larger than Shannon entropy.  The day bits are imprinted on molecules,
rather than in chips, the two entropies will give the same number. The
connection between information and entropy turns out to be very useful
and important in black hole theory, see, e.g.,
\cite{bekensteininformation}.

To try to understand the meaning of the black hole entropy given in
equation (\ref{entropyarealawgeometrical}), one can use the various
formulas for the entropy presented above. But here, for our purposes,
it is simpler if we explore Boltzmann's formula (\ref{boltzentropy0}).
It chiefly claims that one way to think about entropy is that it
is a measure, a logarithmic measure, of the number of accessible
microstates that the isolated system has. Any system, including a
black hole, should follow this rule.  For black holes, there is a
snag, we do not really know what those microstates are, so we cannot
count them to take the entropy. There are several ideas.  One idea is
that the microstates could be associated to the singularity inside the
event horizon, where the crushed matter and the demolished spacetime
lie altogether.  As in the ordinary matter case, one could think that
rearranging these states, somehow lying on the singularity, do not
affect the mass $M$ of the black hole (and $Q$, and $J$ for the other
hairs, if there are those). There are problems with this
interpretation for the entropy. The singularity is in principle
spacelike, in addition it is certainly causally disconnected to the
outside of the black hole, and therefore it is hard to imagine how it
could influence any quantity exterior to itself, let alone to the
exterior of event horizon. This interpretation is related to the
interpretation that the degrees of freedom, are in some measure of the
volume inside the horizon (see, e.g.,
\cite{bekensteinreviewhistory,jacobson}). Moreover, such type of
interpretations are very difficult to implement since no one knows
really what goes on inside let alone in a a singularity, only with a
fully developed theory of quantum gravity can one attempt to
understand singularities.  Another place where the microstates might
be located is in the vicinity of the event horizon area as has been
suggested many times (see, e.g., \cite{wheelerjourneybook} for a
heuristic account, \cite{carlip99,solodukhin,diaslemos} for a
particular implementation, and \cite{lemosreview} for a review).  The
idea beyond this suggestion is that for photons emitted near the
horizon, only those with very high energies, indeed trans-Planckian
energies, can arrive with some finite nonzero energy at infinity, and
so, these photons probe near-horizon Planckian structures, i.e., probe
quantum gravity.  Indeed, light sent from the very vicinity of the
horizon has to climb up the huge gravitational field, or if one
prefers, the huge spacetime barrier set up by the black hole.  In turn
this means that the pulse of light an observer a distance away form
the horizon receives has a much lower frequency (much higher
wavelength) than the very high pulse frequency (very low wavelength)
of the emitted pulse. This is the redshift effect.  The nearer the
horizon the pulse is emitted the higher the effect.  Since in quantum
mechanics frequency and energy are the same thing, $E=\hbar\omega$,
the closer to the horizon the photon is emitted, the more energy it
must get rid off as it travels towards the observer. In effect, there
is an exponential gravitational redshift near the horizon so that the
outgoing photons and other Hawking radiation particles originate from
modes with extremely large, trans-Planckian, energies. But now this is
very important, photons with very high energy, very low wavelengths,
probe very small regions. So the Planckian and trans-Planckian
photons, that arrive at the observer somehow come from regions of
space and time that are themselves quantum gravity regions.  There are
various possibilities for these regions, such spacetime regions may be
discrete, or may be fluctuating in a quantum foam structure, or
whatever. Thus, if one can observe Hawking photons originating from
very close to the horizon of a black hole, one is possibly seeing the
quantum structure of the spacetime. In the context we are discussing,
this means that the entropy should be a feature of the horizon region
itself.  Near the horizon quantum gravity and matter fields are being
probed, and these, alone or together, can be the degrees of freedom
one is looking for to generate the entropy of the black hole.  This
fact led thus to the proposal that the entropy is in the horizon area.
This proposal is very interesting and may solve the degrees of
freedom, or the entropy, problem.  But this follow up from black hole
thermodynamics is not our main concern here, see \cite{lemosreview}
and references therein for more on that.  We have commented on it
solely to get a preliminary understanding of black hole entropy. Even
without understanding where are those degrees of freedom that make up
the black hole entropy one can derive some new consequences, such as
the entropy bounds and the holographic principle.

\section{An Entropy bound involving black holes}

The generalized second law allows us to set bounds on the the entropy
of a given system. Or, in terms of information, it sets bound on the
information capacity any isolated physical system can have. Since this
law involves gravitation, and gravitation together with quantum mechanics 
should provide a fundamental theory, the bound refers to
the maximum entropy up to the ultimate level of description, a given
region can have.

To obtain the bound let us think of the formation of a black hole from
the collapse of some ordinary matter. It is interesting to consider
thus an initial configuration, a star say, and a final configuration,
a black hole.  Consider then any approximately spherical isolated
matter that is not itself a black hole, and that fits inside a closed
surface of area $A$. If the mass can collapse to a black hole, the
black hole will end up with a horizon area smaller than $A$, i.e.,
$A_{\rm bh}\leq A$.  The black hole entropy, $S_{\rm bh}=A_{\rm
bh}/4$, is therefore smaller than $A/4$. According to the generalized
second law, the entropy of the system cannot decrease. Therefore, the
initial entropy of the matter system, $S_{\rm initial}^{\rm system}$,
cannot be larger than $A_{\rm bh}/4$, and so not larger than $A/4$. It
follows that the entropy of an isolated physical system with boundary
area $A$ is necessary less than $A/4$, i.e., $S_{\rm initial}^{\rm
system}<A/4$. So, following ideas devised early by Bekenstein
\cite{originalbekensteinbound}, Susskind \cite{susskindsphericalbound}
through such a simple argument developed this spherical bound. Putting
$S_{\rm initial}^{\rm system}\equiv S$, to clarify the notation,
Susskind's bound reads
\begin{equation} 
S\leq\frac14\,A\,.
\label{entropybound} 
\end{equation} 
One can now anticipate a result which will be further discussed in the
next section: since $A$ is the number of Planck unit areas that tile
the area $A$, the bound says that the number of quantum degrees of
freedom, or the logarithm of the number of quantum states, of the
system within an area $A$ is necessarily equal or less than one
quarter of the number of Planck unit areas that fit in the area $A$.
In brief, following \cite{susskindsphericalbound} the generalized
second law implies the bound (\ref{entropybound}), usually called the
spherical holographic bound for reasons we will see below.

One example one can give that certainly satisfies the bound refers to
two black holes in a box.  Let us put two Schwarzschild black holes of
masses $M_1$ and $M_2$ in a box. The entropy is $S=\frac14 \left(
A_{\rm bh1}+A_{\rm bh2}\right)$. Since $A_{\rm bh1} =4\pi 
R_{\rm bh1}^2= 16\pi
M_1^2$ and $A_{\rm bh2} =4\pi R_{\rm bh2}^2= 16\pi M_2^2$, one has $S=4\pi
\left(M_1^2+M_2^2\right)$.  Now, from a distance, the system should
not be a large black hole of mass $M=M_1+M_2$, otherwise the argument
is of no interest. So, there is a radius for the box $R$, with the
associated area $A$, which obeys $A_{\rm bh1}+A_{\rm bh}<A$, i.e.,
$\frac14\left( A_{\rm bh1}+A_{\rm bh2}\right)<\frac14 A$.  Finally, since
$S=\frac14\left( A_{\rm bh1}+A_{\rm bh1}\right)$ one has $S<\frac14
A$. The bound is clearly satisfied, and it is saturated only when the
box is a black hole. There are many other examples one can think of.

Now this bound suffers from some drawbacks, it only applies to systems
which are initially nearly spherically symmetric, and not much
strongly time dependent. In addition is not a covariant bound, and in
general relativity, all statements should some way or another be put
in a covariant form. These problems were cured by Bousso
\cite{bousso1,bousso2,bousso3}, who has managed to formulate a
covariant entropy bound, (see also
\cite{flanagan,gaolemos1,gaolemos2,gaolemos3}).  Susskind's bound is a
particular case of this Bousso's covariant bound.  In addition, the
original Bekenstein bound \cite{originalbekensteinbound} can be
derived through Susskind bound, and surely, through Bousso's bound
(see \cite{boussoderivationofbounds}). The covariant entropy bound is
fascinating and has been proved correct in very many instances.
However, due to its simplicity, it is useful to stick to the spherical
bound given in equation (\ref{entropybound}). This will take us more 
directly to the holographic principle.

\section{Holographic principle}

\subsection{Definition}

What are the ultimate degrees of freedom, what are the degrees of
freedom of quantum gravity, what are the fundamental constituents of
spacetime?  Portions of ordinary matter are made of molecules, which
are made of atoms, which are made of electrons and nuclei, which
nuclei are made of protons and neutrons, which are made of quarks, and
so on including all known interactions and their associated particles,
up to the quantum gravity level, the fundamental spacetime level. For
ordinary matter and the corresponding cascade of constituents and
interactions we know what and where the degrees of freedom are. For
spacetime we do not, yet, unfortunately. However, even without knowing
of what the spacetime is made of we can extract limits for the number
of such degrees of freedom, and other relevant information from the
entropy bounds discussed previously.

Indeed, based on his own ideas about entropy bounds and even before
the spherical bound was advanced, 't Hooft
\cite{thooftonholoprinciple} proposed that the degrees of freedom of a
region of space circumscribed within an area $A$ are in the area
itself. This is counter to the results of everyday physics which give
that the entropy is proportional to the volume of the region, and so
the degrees of freedom of these usual systems are in this sense in the
volume. A usual system has entropy, or information, inside it. For
instance, in a book, the information is contained inside (in the
volume), not in the cover (in the area). One knows that reading the
title of a book is not enough at all to know what is inside. This also
happens for all usual thermodynamic physical systems upon a
statistical physics treatment. However, there are system that do not
follow this rule. These are the black holes, which once more reveal
themselves as the most fundamental objects to uncover the secrets of
nature. For black holes, the entropy and information of what is inside
is projected in the area. Since black holes are of fundamental
importance, both in gravitation and quantum theory, they are the ones
that dictate the ultimate rule that should be obeyed. Thus, the
generalized second law, derived from black hole physics, together with
the bounds above conjure to give the result that the degrees of
freedom are in the area itself.

To be definitive, define first the number of degrees of freedom
$N_{\rm f}$ of a quantum system as the logarithm of the number of
quantum states $\Omega$ of the system, with $\Omega$ being the same as
the dimension of the Hilbert space of the system, (parts of this
exposition follows
\cite{susskindsphericalbound,bousso3,thooftonholoprinciple}).  So
$N_{\rm f}=\ln \Omega$.  This generalizes the idea of degrees of
freedom of a classical system. To have an idea of what $N_{\rm f}$
means, take, for example, a spin system with 1000 spins, each spin
being able to be up or down only. Such a system 
should have around 1000 degrees of
freedom. In fact, from the definition above, since there are two
states for each spin, one has that the number of states for the whole
system is ${\cal N}=2^{1000}$.  So its number of degrees of freedom is
$N_{\rm f}=1000\ln 2$. In terms of information, following Shannon,
this means that the system can store 1000 bits, or its Shannon entropy
is 1000, as mentioned in Section III. This system is small,
thermodynamics systems are huge in comparison. For a given isolated
thermodynamic system, with entropy $S$, the number of independent
quantum states is $\Omega={\rm e}^S$, in natural units, see equation
(\ref{boltzentropy0}).  So, for a thermodynamic system, $N_{\rm f}$ is
related to the entropy $S$, in fact, following the definition above,
they are the same in natural units, $S=N_{\rm f}$.  For instance, in
order to see that such a definition is reasonable, recall that the
number of states $\Omega$ of an ideal gas with fixed energy $E$,
volume $V$, and number of particles $N$ can be written as
$\Omega=\left[{\rm e}^{5/2}\,(V/N)(4\pi mE/3N)^{3/2}/h^3\right]^N$, so
that since $V=L^3$, where $L$ is the dimension of the enclosure say,
and $(2mE)^{1/2}=\bar p$, where $\bar p$ is a typical momentum of the
particles, one finds $N_{\rm f}=6N\, \alpha$ where $\alpha$ is a
number of the order one or so, proportional to a logarithm term. Thus,
$N_{\rm f}$ as defined gives roughly the classical number $6N$ of
degrees of freedom as expected.  Of course, through this definition
one has exactly $S=N_{\rm f}= 6N\left[{5/12}+(1/6)\ln\left((V/N)(4\pi
mE/3N)^{3/2}/h^3\right)\right]$, which is the Sackur-Tetrode formula
(see, e.g., \cite{cowan}).

Let us suppose then that we are given a nearly spherical finite region
of volume $V$ with boundary area $A$. Suppose again that, initially,
gravitation is not strong enough, so that spacetime is not time
dependent and all the relevant physical quantities are well
defined. One can consider then that the nearly spherical region has
some matter content.  However, this content ultimately does not
interest us, we can forget about the solid, liquid, gas, or vacuum
that fills up the region.  At the ultimate level one is only
interested in the region itself, in the spacetime itself alone. One
wants to know what are the states themselves of that region and what
is their number, at the most fundamental level.  So we want to know
how many degrees of freedom are there for the fundamental system, or
how much complexity there is at the fundamental level, or how much
information one needs to specify the region. One way to start out and
see where it leads to is to pick up a theory that has given fruitful
results in ultra microscopic physics. This theory is quantum field
theory.  It works extremely well in flat spacetime, and with care it
can be extrapolated to curved spacetime \cite{birrel}.  A quantum
field is described by harmonic oscillators at every point in
spacetime.  A quantum harmonic oscillator has an infinite number of
states and so an infinite number of quantum degrees of freedom.  So,
there are infinite number of degrees of freedom at every spacetime
point in a quantum field. Moreover, within a volume $V$ there are
infinite number of points. Thus, a quantum field in a given spacetime
background has, by this rationale, a huge infinite of infinite number
of degrees of freedom. So it seems.  However, one can easily argue,
that the number is indeed huge, albeit finite.  Indeed, gravity
together with quantum theory show that there is a minimum length scale
given by the Planck length $l_{\rm pl}$, and a maximum energy scale,
the Planck mass $m_{\rm pl}$, beyond which any theory of distances and
scales does not make sense. So, crudely, one might guess that
there is one oscillator per $l_{\rm pl}$, each with maximum energy
$m_{\rm pl}$ (more energy than this turns spacetime into a black
hole). One can now think that each spacetime volume $V$ has $V/V_{\rm
pl}$ oscillators and each oscillator has a finite number of states $n$
say, which is large, (the highest energy state for each oscillator is
given by the Planck energy).  So, in Planck or natural units, the
total number of states is $\Omega\simeq n^V$ and thus the number of
degrees of freedom is $N_{\rm f}\simeq V\ln n$, i.e., $S\simeq V\ln n$
(see \cite{bousso3} for more details).  So, if this conclusion is
fully correct, a fundamental theory needs to account for an entropy
proportional to the volume or bulk of each region being considered,
i.e., the disorder of the region goes with the volume.

But this na\"ive reasoning fails when gravity is included. The
fundamental theory has to include gravity, for sure, and when this is
done one finds that a fundamental theory needs only to account for an
entropy proportional to the surface area, and this is much less than
entropy proportional to the volume.  Let us see this in more
detail. Entropy is a measure of the logarithm of the number of
microstates of a given system macroscopically specified, so that, as
seen, entropy is also a measure of the number of degrees of freedom of
the system. Now we know that given an area $A$ there is a bound for
the entropy $S\leq \frac14 A$ in Planck units (i.e., $\frac{S}{k_{\rm
B}}\leq \frac14 \frac{A}{A_{\rm pl}}$ restoring units) for any
system. Any system, including the fundamental system, has to obey this
bound. When we have an adequate quantum gravity it will give an
entropy for the quantum system which is equal or lower than this
bound. Now a black hole with this same area $A$ saturates the bound,
so one can say there are systems that saturate the bound. Thus the
number of degrees of freedom of a sphere of area $A$, and the 
related number of states are given by, respectively, 
\begin{equation} 
N_{\rm f}=\frac14\,A\,,\quad {\rm and} \quad\Omega={\rm e}^{\frac14A}\,.
\label{numerofsatesfundamental} 
\end{equation} 
That the number of states has to be given by
(\ref{numerofsatesfundamental}) can be argued more effectively using
unitarity, which claims that an initial state evolves in a well
defined manner to a final state, such that probability in quantum
theory is conserved.  Essentially, it says one can derive the final
state from the initial and vice-versa.  Given an initial object, or
region, suppose that the number of states of the Hilbert space for it
goes roughly with ${\rm e}^V$.  Allow the object, or the region, to
evolve into a black hole of the same size of the region.  
Then the new number of states is ${\rm
e}^{A/4}$, where $A$ is the area enclosing $V$.  But this number is
much less than the initial one, so one cannot recover the initial
state from the final one, the states would not evolve unitarily. Thus,
one should start with ${\rm e}^{A/4}$ as the initial number of states.

Now, the number given in (\ref{numerofsatesfundamental}) is much
smaller than the number $n^V$ guessed earlier, for lengths larger than
about the Planck size.  One can understand this much lower number than
the one given by the na\"ive guess of quantum field theory, by
invoking heuristic arguments coming from the inclusion of gravity (see
again \cite{bousso3} for more details). It is true we have imposed,
na\"ively, that there is at most one Planck mass per Planck volume. So
there is a high energy cut off, and modes with higher energy than that
do not exist and do not contribute to the entropy. That is fine.  But
this cut off at large scales, scales larger than Planck scales, gives
that, within a region of radius $R$ and assuming roughly a constant
field density, the mass can scale as $M/M_{\rm Pl}\sim \left(R/R_{\rm
Pl}\right)^3$, i.e., $M\sim R^3$ in Planck units. This cannot be
right, since we know that for sure $M/M_{\rm Pl}\buildrel<\over\sim
R/R_{\rm Pl}$, i.e., $M\buildrel<\over\sim R$ in Planck units. For
$M\buildrel>\over\sim R$ one forms a black hole, the most massive
object that can be localized in the sphere of radius $R$. Thus at face
value it seems that one should rather assume that the field content
(gravity and possibly other fields) density goes at most as $1/R^2$
rather than constant.  And so there are many less states than
na\"ively one could guess.  Due to gravity, a long range universal
field, the energy of the field content is lower in large volumes than it
could possibly be in small Planckian volumes.  Thus crudely, the
entropy, which in many ways is related to the energy, is also
drastically reduced. The conclusion is that na\"ive field theory seems
to yield more degrees of freedom than those that can be used for
generating entropy, or to store information.  So, there are at most
$A/4$ degrees of freedom inside a region whose volume is surrounded by
an area $A$. Most systems have less than $A/4$ degrees of freedom such
as any system made of ordinary matter. One system that strictly
matches the bound is a black hole, which has precisely $A/4$ degrees
of freedom.  If one has any system, and wants to excite more degrees
of freedom than those given by the bound, then one forms a black
hole. A black hole is an object that has the maximum entropy for an
outside observer.  Perhaps there can be other objects with such an
entropy, e.g., quasi black holes 
(see \cite{weinberglue,lemoszaslavskii}), but not
object has larger entropy. Summing up what we have seen so far, we can
say that Bekenstein's and Hawking's works, coupled to the Susskind
bound, states that a fundamental theory, one in which gravity is
included, has a number of degrees of freedom proportional to the area,
which leads to fewer degrees of freedom, and so less entropy or less
disorder, than the theory would have to have had the entropy of a
region been proportional to the volume, rather than the area.

Then one can go a step further, as 't Hooft did
\cite{thooftonholoprinciple}, actually before the spherical and the
covariant bounds were discussed. If the maximum entropy, obtained from
fundamental degrees of freedom, in a given region of space, is
proportional to the area, rather than the volume, then the degrees of
freedom should lie in the area of the region.  This is the basis of
the holographic principle.  It states: a region with boundary area $A$
is totally described by at most $A/4$ degrees of freedom (in Planck or
natural units), i.e., about one bit of information per Planck area.
In a sense, the description of the processes that happen within the
region's volume, is projected into the surface of that region, in the
same way as the visual perception of a three dimensional region can be
encoded in a hologram, a two dimensional sheet.  So, in principle,
there are two possible descriptions, the volumetric or three
dimensional, and the areal or two dimensional description, the latter
one being certainly more economical.  In order to grasp better this
idea let us introduce the following allegory \cite{greene}.  Imagine
that a futuristic plane is surrounded by a hypothetical giant two
dimensional spherical screen located in space.  And that all the
activities and happenings on such a planet, through illumination, are
projected onto this screen. The image on the giant screen would be a
blow down of the three dimensional world to two dimensions.  If the
projection is accurate enough, there are two sorts of people, one two
dimensional, the other three dimensional.  But in such an accurate
case, people in both scenarios can think of themselves as equally
alive, and the other as the mirage, each containing the same amount of
information and each being described through equivalent mathematical
theories, with no theory being more correct than the other.

\subsection{Implementation}

Concrete examples which satisfy the holographic principle have been
found in anti-de Sitter spacetimes, i.e., spacetimes with a negative
cosmological constant. A de Sitter universe is one with a positive
cosmological constant that creates a universe uniformly accelerating,
and present observational results indicate we leave in such a
universe. On the other hand, a negative cosmological constant has the
property of giving a uniform gravitational attraction over all of the
spacetime. Such a spacetime has uniform negative curvature, and due to
the relentless constant attraction this spacetime has a boundary, it
is as if the spacetime is set up in a box with some definite length.
For instance, in this spacetime, a massless particle can travel in a
finite time from any point in the interior to spatial infinity and
back again.

Anti-de Sitter spacetimes appear often in string theory, or its
M-theory generalization, as well as in several corresponding low
energy limits that yield supergravity theories.  Now, when string
theory is properly used in an anti-de Sitter spacetime one finds that
it is equivalent to a quantum field theory on the boundary of that
spacetime \cite{maldacenaadscft}.  The first instance in which this
holographic result was found is not directly applicable to our real
universe, first because the cosmological constant of the universe is
positive rather than negative, and second because in
\cite{maldacenaadscft} it was found that the calculations simplify
if one works in a five dimensional (four space and one
time dimensions) anti-de Sitter spacetime, ${\rm AdS}_5$, rather than
the more usual four dimensional one, ${\rm AdS}_4$. String theory is
well formulated only in ten dimensions, so to be precise, one can also
include compactified dimensions. In fact, the example is given in the
context of ${\rm AdS}_5$ times the five sphere ${\rm S}_5$, so that
the whole spacetime is ${\rm AdS}_5\times{\rm S}_5$.  The equivalent
boundary quantum field theory arises from the boundary of ${\rm
AdS}_5$.  In this setting, one can argue that the physics experimented
by an observer living in the bulk of the ${\rm AdS}_5$ spacetime can
be completely described in terms of the physics taking place on the
spacetime's boundary.  This initial result, valid for a five
dimensional anti-de Sitter spacetime and its four dimensional boundary
dual, was later exhibited in many other situations and other
dimensions, including the more usual four dimensional anti-de Sitter
spacetime, being in this case dual to a quantum field in three
spacetime dimensions.  Generically, one finds that the bulk and the
boundary descriptions are equivalent, none of the descriptions is more
complete or important than the other. In the bulk description
gravitation operates and spacetime is d dimensional, say, whereas in
the boundary description there is no gravity but a quantum field
theory, conformal in nature, operating in a d$\,$-1 dimensional flat
spacetime. It is as if there is a duality between this d dimensional
spacetime and its d$\,$-1 dimensional boundary.  It then means that
two different theories, acting in spacetimes with different
dimensions, are equivalent. This reinforces the idea that beings
living in the d dimensional spacetime would be mathematically
equivalent to beings living in the d$\,$-1 dimensional one, there is
no way to distinguish between them. This is certainly an interesting
implementation of the holographic principle of 't Hooft
\cite{thooftonholoprinciple}.  Technically, it is also fruitful, since
difficult calculations performed on the bulk spacetime can perhaps be
easily done in the quantum field theory on the boundary and vice
versa. For instance, one can show that a black hole in anti-de Sitter
spacetime is equivalent to hot radiation in the boundary, and that the
mysterious black hole entropy is equivalent to the radiation entropy
\cite{wittenbhradiation}.  In addition, it may give insights into the
information problem in black hole physics. We have not yet explicitly
mentioned it.  This problem is related to the entropy interpretation
problem of what and where are the degrees of freedom corresponding to
the black hole entropy. We have argued that the black hole seems to
hide many features inside the horizon. For instance it possibly hides
all the information that the original star had before it collapsed
into a black hole. But now, how can we get information to reobtain
those features?  If the inside of black holes are disconnected to the
outside world, then classically this information seems to have
disappeared. Is there an information loss, and with it a break of
unitarity?  It has not yet been shown that information is lost or not
lost when one throws objects through a black hole, following Hawking's
original suggestion of information loss. But, in this string theory
description, a black hole is now dual to a lower dimensional world in
which it seems information is never lost. So there is hope in solving
this problem.  Another place where it can be of use is in quantum
field theory itself. The reason is that anti-de Sitter spacetime
yields relatively easy calculations, whereas calculations on quantum
fields are technically hard. For instance, one cannot yet derive
the proton and neutron properties from quantum chromodynamics, the
theory of quarks, a well understood theory, but extremely hard to
solve. One can now try to solve these properties using the above
duality.

\section{Conclusions}

We have trodden a long way from the first ideas in black holes with
their associated event horizons all within the context of pure
classical general relativity, passed through semiclassical
calculations meeting the concepts of entropy and temperature for black
holes, and then through the statistical physics connection of entropy
and its associated second law, arriving at the maximum number of
degrees of freedom a fundamental theory, one which includes quantum
gravity, can have. Surprisingly, this number goes with the area $A$ of
the region, rather than with the volume $V$. In turn this means first
that the holographic principle should be valid, i.e., a region with
boundary area $A$ is totally described by at most $A/4$ degrees of
freedom (in natural units), and second we need a fundamental theory
that incorporates this principle. As we have seen, local quantum field
theory is certainly not such a theory.  Taking this idea seriously,
one can advance that the universe can indeed be described by a model
with one less dimension, in the sense, that the formulation of the
fundamental theory can be done in a lower dimension. Perhaps, as
string theory suggests, the fundamental theory can be formulated in
the dimensions we are used to (in our case three plus one spacetime
dimensions), as well as in the holographic dimensions (in our case two
plus one spacetime dimensions). In this case, one should also be able
to find a dictionary, or a map, between both formulations.  This idea
that one can trade spacetime dimensions in a fundamental description,
leads one to speculate that, conceivably, spacetime itself is not a so
fundamental concept. Of course, if true, such considerations are
destined to enter into the philosophy dominion and radically transform
our notions of what space and time are. Is the universe a hologram?
Is there a shadow universe in which our bodies exist in a compressed
two dimensional form?  The answer lies ahead.

\vskip 1cm
\section*{Acknowledgments}
I thank the Organizing Committee of the meeting {\it Advances in
Physical Sciences}, held at the Universidade de Aveiro, September
2005, for the invitation to lecture on this topic, and in particular,
I thank Luis Carlos for his patience with the manuscript.  I thank
Ant\'onio Luciano Videira, to whom this meeting is devoted, for
initiating me in scientific matters, namely black hole thermodynamics
back in 1980.

\end{document}